\documentstyle[12pt]{article}

\newcounter{defin}

\newcommand\re[1]{{(\ref{#1})}}
\newtheorem{dfn}[defin]{Definition}

\newcommand{\be}{\begin{equation}}
\newcommand{\ee}{\end{equation}}
\newcommand{\bea}{\begin{equationarray}}
\newcommand{\eea}{\end{equationarray}}
\newcommand{\ba}{\begin{array}{c}}
\newcommand{\ea}{\end{array}}

\newcommand{\al}{\alpha}
\newcommand{\e}{\epsilon}
\newcommand{\ve}{\varepsilon}
\newcommand{\vp}{\varphi}
\newcommand{\pa}{\partial}
\newcommand{\fr}{\frac{1}{2}}

\newcommand{\la}{\lambda}

\begin{document}

\thispagestyle{empty}

\begin{flushright}
\vspace{1mm} hep-th/0008105\\
FIAN/TD/12/00\\
August 2000\\
\end{flushright}

\vspace{10mm}

\begin{center}
{\bf  Point particle in general
background fields \\ and generalized equivalence principle$^\diamond$} \\

\vspace{10mm}
Arkady Yu. Segal ${}^{(1)}$
\\

\vspace{1cm}   {\it
I.E.Tamm Department of Theoretical Physics, Lebedev Physics
Institute,\\
Leninsky prospect 53, 117924, Moscow, Russia\\
e-mail: segal@lpi.ru  }

\vspace{2mm}

\end{center}
\vspace{5mm}
\begin{abstract}
The model of point particle in general external fields is considered
and the generalized equivalence principle is suggested identifying
all backgrounds which give rise to equivalent particle dynamics.
The equivalence transformations
for external fields are interpreted as gauge ones.
The gauge group appears to be a semidirect product
of all phase space canonical
transformations to an abelian ideal of "hyperWeyl" transformations
and includes  $U(1)$ and general coordinate symmetries as a subgroup.
The implications of this gauge symmetry are considered and a
connection of general backgrounds to the
infinite collection of Fronsdal gauge fields
is studied. Although the result is negative and no direct
connection arises, it is discussed how
higher spin fields could be found among general external fields
if one relaxes somehow the equivalence principle. Besides,
the particle action in general backgrounds is shown to reproduce
the De Wit-Freedman point particle -- symmetric tensors
first order interaction suggested
many years ago, and generalizes their result
to all orders in interaction.
\end{abstract}
\vspace{1cm}

--------------------------------------

{\tt $(1)$ on leave of absence from}

Department of Physics, Tomsk State University, Russia.

\vspace{5mm}
$^\diamond${\it The materials of this paper were
presented by the author at the
International Conference
"Quantization, Gauge Theory and Strings",
dedicated to the memory of Professor E.S. Fradkin,
Moscow, June 5-10th, 2000.}

\newpage
\section{Introduction.}

Efim Samoilovich Fradkin was a brilliant scientist who
had been always interested in the most important directions of theoretical
physics and made many significant contributions to modern
high energy physics.

One of the topics he studied in depth is the
{\it higher spin problem} which presents itself the
issue of constructing consistent interacting lagrangian theories of
massless fields of arbitrary spin and, particularly,
of coupling the arbitrary spin massless fields to gravity.

Generally speaking, the free higher spin models are gauge theories
constructed in terms of symmetric tensors,
so the higher spin problem is intimately related to the
construction of the interacting theory of symmetric tensor gauge fields
(including gravity).
In the present  paper, we try to put forward a new approach to the problem
which uses the covariance of point particle -- external fields interaction
to provide the full nonlinearized gauge transformations for
exterior fields being a collection of symmetric tensors
$H =\{ H(x), H^{m}(x), H^{m_1m_2}(x),...,H^{m_1...m_s}(x),...\}$ (every rank appears once),
the low 0,1,2-rank tensors correspond to low spin dilaton, electromagnetic
and gravitational fields.
Despite the overall result will be negative in the sense
the gauge transformations appear to be too restrictive
to associate $H$'s with higher spin gauge fields, our
study provides the hope the connection to higher spin  fields may be
found after some modifications of gauge transformations or fields content.

Before passing to the main text we recall basic
features of of higher spin theories and provide a brief historical
survey. This will give us additional grounds in favour of our program.
(The reader may skip this part of the text and
pass directly to Sec. \re{classica} where the program starts).

The lagrangians models of free
arbitrary spin massless fields were constructed
by Fronsdal and Fang and Fronsdal \cite{fronsdal1}  in $4D$
Minkowski and anti-de Sitter spaces,
and then reformulated in the "gauge form" by
Vasiliev \cite{Vasiliev:1980as} and, independently, by
Aragone and Deser \cite{Aragone:1980rk} for fermion case.
Later on, the
actions for arbitrary spin massless fields in flat $d$-dimensional space
were built by Labastida \cite{lab}, and
for symmetric Young tableaux bosons in $d$-dimensional $AdS$ space
by Lopatin and Vasiliev \cite{Lopatin:1988hz} and for fermions by
Vasiliev \cite{Vasiliev:1988tk}.
E.g., the theories of totally symmetric spin-$s$ boson fields in
Minkowski space of any dimensions are formulated
in terms of symmetric rank-$s$ tensor fields
$\vp_{m_1...m_s}(x)$
satisfying the {\it double-tracelessness} constraint
\be \label{dtrace}
{{\vp_{kl}}^{kl}}_{m_5...m_s} =0 ,
\ee
where the contraction is performed by the background metric.
For $s=0,1,2,3$ this constraint is satisfied automatically.

For $s\geq 1$, the action is unambiguously determined by the requirement
of absence of higher derivatives and invariance w.r.t. gauge
transformations (the round brackets stand for symmetrization)
\be \label{frlaw}
\delta \vp_{m_1...m_s}     =\partial_{(m_1} \ve_{m_2...m_s)}
\ee
whit {\it traceless} gauge parameter:
\be \label{etrace}
{\ve^{k}}_{km_3...m_{s-1}} =0.
\ee
For $s=1,2$, this constraint is automatically fulfilled.

The invariant action reads
\be  \label{fraction}
\begin{array}{cllc}
{\cal A}_s[\vp_s]=\fr (-)^s \int d^d x \bigl\{
\pa_{n} \vp_{m_1...m_s} \;\pa^{n}  \vp^{m_1...m_s}& &\\ & &\\
-\fr s(s-1) \pa_{n} {\vp^{k}}_{km_1...m_{s-2}}\; \pa^{n}{{\vp^{k}}_k}^
{m_1...m_{s-2}} & &\\ & &\\
+ s(s-1) \pa_{n} {\vp^{k}}_{km_1...m_{s-2}}\; \pa_{l} \vp^{nlm_1...m_{s-2}}
-s \,\pa_{n} {\vp^{n}}_{m_1...m_{s-1}}\; \pa_{k} \vp^{km_1...m_{s-1}}& & \\ & &\\
-\frac{1}{4} s(s-1)(s-2)
\pa_{n} {{\vp^{k}}_k}_{nm_1...m_{s-3}}\; \pa_{r}
{{\vp^{l}}_l}^{rm_1...m_{s-3}}\bigr\} .& &
\end{array}
\ee
For $s=0$
there are no gauge parameters at all, i.e. a theory is not a gauged one.
Nevertheless, the formulas \re{fraction} make sense in this case
and provide the massless scalar theory.
Though the actions \re{fraction} were originally constructed  in $4D$
\cite{fronsdal1} they do not contain explicitly the dimension $d$ of
spacetime and describe consistently massless higher spin dynamics for any $d$.
We shall refer to them as to Fronsdal actions.

It is not easy to preserve the gauge symmetries \re{frlaw} upon
introducing interaction. Particularly, already the minimal
coupling to gravity is inconsistent as it breaks the linearized gauge
invariance by terms proportional to the full Riemann curvature:
$
\delta {\cal A}_s \sim \int d^d  x \vp R \ve
$
which could not be cancelled by contributions from any nonminimal terms.
This inconsistency is a core of the higher spin problem
\cite{ardeser}, \cite{Weinberg:1980kq}.

The way out was found by Fradkin and Vasiliev who noticed that
a good starting point for perturbative analysis is anti-de Sitter vacuum
for metric instead of Minkowski one \cite{vas1}.
It became possible
to avoid the higher spin "no-go" restrictions at least in the first
nontrivial order in interaction.
Furthermore, the all-orders interaction had been conjectured to be consistent
only provided all higher spin fields are used altogether.

A few years later, having invented the powerful {\it unfolded formulation}
technology Vasiliev had constructed consistent nonlinear
{\it equations of motion} describing a supersymmetric interacting system of
higher spin massless fields (every spin appears twice) in
$4D$ space-time \cite{vas2}. The equations of motion of
interacting $3D$
higher spin - matter system were constructed \cite{vas234}, \cite{vas3} and
have been shown to admit (for the massless matter case)
a nontrivial action principle \cite{vas3s}.
Besides, $2D$ actions for higher spin interactions of matter fields
have been given \cite{vas234}, \cite{vas2d}.
The reviews of this activity may be found in \cite{vasO}.
One may pick up the main properties of higher spin theories
that seem to persist in any model describing consistent
interactions of higher spin massless fields:

1) the model is invariant under infinite-dimensional gauge symmetry
which includes in particular Yang-Mills and general coordinate transformations;

2) there is an infinite number of massless fields of all spins
including low spin dilaton, Yang-Mills and gravity fluctuations;

3) a natural background for metric is a nonzero curvature space
rather than flat one; the straightforward flat limit is singular as interaction
contains terms proportional to the inverse powers of cosmological constant;

4) the theory is unitary at the quadratic level and the linearized
equations of motion do not contain higher derivatives,
but interaction necessarily involves higher derivatives, so that
a full theory may even be nonlocal.

By now, the higher spin problem is elaborated by many authors and
there is a lot of interesting results \cite{vasO},\cite{bmv},\cite{vas5},
\cite{metsaev1},\cite{metsaev2},\cite{ks},\cite{gks},\cite{sibir},\cite{klishevich},
\cite{pashnev}, \cite{Bellon:1987ki},
\cite{aragdes},\cite{nograv},\cite{Sezgin:1998gg},
\cite{Marnelius:1990be},\cite{plus},\cite{ners}\cite{Bandos:1999rp}.
It is likely  that
a desicive solution of higher spin problem in any dimension which
should provide a consistent lagrangian description of higher spin interactions
may exist but it still requires a lot of study.
If such a fundamental theory
exists indeed, much probably it is based on deep and simple geometrical
principles associated with its huge gauge symmetry group.
It is tempting to  anticipate these principles may appear to be
as clear as gauge principles of Einstein or Yang-Mills theories and
determine the action functional unambiguously up to a well controlled
arbitrariness and fields redefinition.

One has another hint of existence of higher spin theories.
The off-shell fields content, the gauge transformations (\ref{frlaw})
and the  form of actions
(\ref{fraction})  are the same
in any dimension. This fact can be thought about as a manifestation of
a geometry which can be developed uniformly in any dimension,
this hypothetical geometry has to provide the
invariant actions for arbitrary spin fields which after the linearization
over some vacuum solution give Fronsdal actions, just like $U(1)$ geometry
gives the Maxwell action, and like Riemann geometry underlies
the Einstein action.

The basic input governing  Einstein gravity
is the equivalence principle, similarly, the
electrodynamics action may be considered as a realization of the "$U(1)$
equivalence principle".
{\bf In this paper,} we propose a higher-spin extension of the equivalence
principle, which unifies $U(1)$ invariance, general-coordinate
invariance and a huge amount of another gauge symmetries
into one simple
gauge law associated with the {\it geometry of point particle
phase space}.

We study the classical point particle in general external fields,
and postulate the  {\it generalized equivalence principle}:
two external fields configurations are equivalent if
the particle dynamics
in these configurations are equivalent (for the careful definition
of equivalent particle dynamics, see the main text),
it gives rise to nonlinear, {\it background independent}
gauge transformations, which contain
$U(1)$ and general coordinate transformations subgroups, and of course
contain much more. On the other hand, general background $H$
is given by the infinite set of symmetric tensors, every rank appears once
$H =\{ H(x), H^{m}(x), H^{m_1m_2}(x),...,H^{m_1...m_s}(x),...\}$
the low 0,1,2-rank tensors correspond to low spin dilaton, electromagnetic
and gravitational fields. For higher rank fields,
the gauge invariance appears to be {\it too restrictive} to associate them
with Fronsdal gauge fields.
We analyze why it happens by studying the linearization of gauge
transformations implied by the generalized equivalence principle
around natural vacuum solution corresponding to the flat metric
and constant dilaton and discuss some modifications which may
cure the situation.

As a byproduct of our approach it appears possible to
reproduce an old result by De Wit and Freedman \cite{DeWitFreedman}
on the point particle -- symmetric tensors first-order interaction
and generalize it to all orders.

In Conclusion we briefly discuss the results and some perspectives.

\section{Point particle in general external fields and the generalized
equivalence principle.}  \label{classica}

The basic observations come from the consideration of the dynamics of
ordinary point particle in $d$-dimensional spacetime ${\cal M}^d$. The
worldlines $x^m(\tau)\,,\,m=0,...,d-1$ describe evolution of the particle.
Clearly, the worldlines differing by $\tau$-reparametrization
\be  \label{dif}
x'^m(\tau'(\tau)) = x^m(\tau)
\ee
present equivalent space-time evolution, therefore the particle action
\be   \label{lagr}
S=\int d\tau  L \left(x(\tau), \dot{x}(\tau), x^{(2)} (\tau),...
x^{(n)} (\tau),...\right)
\ee
should be reparametrization invariant.
From now
on we suppose that Lagrangian $L$ does not depend on  higher
$\tau$-derivatives $x^{(2)},...x^{(n)},... \;\;$, then
$L$ should be a homogeneous function of $\dot{x}^m$ of a first order.
The theory admits covariant constrained hamiltonian formulation
\cite{svyazi}, which is built as usual by introducing the
$x^m$-conjugated momenta $p_n$ and making the Legendre transform. As a
consequence of reparametrization symmetry, the model is determined by a
single first class constraint $H(x^m , p_n)\approx 0$.
In summary, the hamiltonian action reads
\be \label {ham}
S_H [x(\tau), p(\tau), \lambda(\tau)]=\int d\tau \{p_m \dot{x}^m -\lambda H
(p, q)\},
\ee
$\lambda$ is a Lagrange multiplier to the unique first
class constraint $H\approx 0$ which we shall call Hamiltonian.

Varying the action (\ref{ham}) w.r.t. $x, p, \lambda$ one
gets the dynamics equivalent to that derived from the Lagrangian, in the form
of Hamilton-like equations with the Hamiltonian $H$ and
the canonical Poisson bracket
\be \label{eomxp}
\ba
\{x^m, p_n\} =\delta^m_n,\\ \\
\dot{x}^m=\lambda\, \{x^m, H\} ,\qquad
\dot{p}_n=\lambda\, \{p_n, H\}
\ea
\ee
plus the condition
\be \label{mshell}
H(x,p)=0.
\ee
The last equation implies that in order to set a nontrivial particle
dynamics $H$ have to possess zeroes as otherwise there are no
solutions for the classical equations of motions.

In the hamiltonian picture, the worldline reparametrization symmetry
(\ref{dif}) becomes the local symmetry of the action (\ref{ham})
\be  \label{invar}
\delta{x^m(\tau)}=\mu(\tau) \{x^m, H\} (\tau), \qquad
\delta{p_m(\tau)}=\mu(\tau) \{p_m, H\} (\tau),\qquad \delta
\lambda=-\dot{\mu}.
\ee
This local symmetry may be fixed by implying the gauge condition
\be \label{gaugecond}
\dot{\la}=0 \;\;\Rightarrow \ddot{\mu}=0
\ee
after that only global modes of gauge transformations survive
$\mu(\tau) = \nu_1 \tau + \nu_2$.

In fact, the hamiltonian action (\ref{ham}) with general
Hamiltonian $H(x,p)$ describes
{\it all possible scalar particle dynamics}  on ${\cal M}^d$
with Lagrangians independent on higher $\tau$-derivatives.
A natural question is  which Hamiltonians are {\it equivalent} in the sense
they set {\it equivalent particle dynamics}.
To answer the question,  first of all
it is necessary to specify which particle dynamics are considered as
equivalent ones.  This topic is rather subtle, e.g. it is well known  any
finite dimensional hamiltonian dynamics may be represented locally in
appropriate action-angle coordinates as $\dot{X}^m(x,p,\tau) = A_n\;,\;
\dot{P}_n(x,p,\tau)=0$, where $X^m(x,p,\tau),P_n(x,p,\tau)$ is some
$\tau$-dependent canonical transformation. Thus, all finite dimensional
models with the same dimension seem to be locally isomorphic.
Our definition ({\it the generalized equivalence principle})
is:
\begin{dfn} \label{def}
\noindent Two particle dynamics determined by the
actions (\ref{ham}) with Hamiltonians $H'$ and $H$ are called equivalent
if there exists a $\tau$-local, $\tau$-independent,
continuous change of variables
$x'(x,p,\lambda)$, $p'(x,p,\lambda)$, $\lambda'(x,p,\lambda)$ such
that the actions are equal up to an integral of a total $\tau$-derivative:
\be \label{Equivalence}
S_{H'}[x',p',\lambda']=S_H[x,p,\lambda]
\ee
In this case, the Hamiltonians $H'$ and $H$ are called equivalent: $H' \sim H$.
\end{dfn}
Let us analyze the definition. First, it is clear that Hamiltonians
differing by a canonical transformation are equivalent:
\be \label{canon}
\begin{array}{c}
\bigl(\mbox{the transform} \left(x'(x,p),p'(x,p)\right) \mbox{ is
canonical} \;\;,\;\;\lambda'=\lambda \;\;,\;\;\;  H'(x',p') = H(x,p)\bigr)
\Rightarrow \\ \\ \Rightarrow H' \sim H,
\end{array}
\ee

since the kinetic term $\int d \tau p_m \dot{x}^m$ is invariant.

Another equivalence of Hamiltonians comes from the transformation
\be  \label{weyl}
\ba
x'^m=x^m\;,\;p'_n=p_n\;\;,\;\;\lambda'=A^{-1}(x,p) \lambda \;\;,\;\;
H'= A(x,p) H,\\ \\ \Rightarrow H' \sim H,
\ea
\ee
where $A(x,p)$ is a nonzero function on the phase space. We call this
equivalence hyperWeyl transformations, while ordinary Weyl dilations
are associated with $p$-independent $A$.

The action of infinitesimal equivalence transformations (\ref{canon},\ref{weyl})
on $H$ may be represented in the form
\be  \label{bs}
\delta H(x,p)  =  \{\epsilon(x,p), H(x,p) \} +  a(x,p) H(x,p)
\ee
where $\epsilon$ is a generating function of the canonical transformations
while $a$ corresponds to the infinitesimal form of (\ref{weyl}) after
the substitution $A =e^a$.
These equivalence transformations form the infinite-dimensional Lie
algebra $\cal G$, being the semidirect product of canonical and
hyperWeyl ones:
\be  \label{ctrs}
\ba
[\delta_{(\e_1,a_1)} ,
\delta_{(\e_2,a_2)} ] H =  \delta_{(\e_3,a_3)}  H\\ \\
\e_3= \{\e_1,\e_2\} \;\;,\;\; a_3=\{\e_1,a_2\} - \{\e_2,a_1\}.
\ea
\ee
The equivalence transformations (\ref{bs}) may seem to be strong
enough to identify all Hamiltonians at all, in this case our definition
could be meaningless.
However, this is not the case.
Formally one could say any two Hamiltonians are connected by some
canonical change of variables \re{canon}. However, in general case
the change is highly singular and can not be considered as an
equivalence transformation.
Below we will return to this issue in more detail by finding
some natural invariant of \re{bs}
(Sec \re{fff}), now we
concentrate on hyperWeyl transformations (\ref{weyl}). It is clear the
Hamiltonians $H_1$ and $H_2$ are not identified by hyperWeyl
transformations (\ref{weyl}) if the surfaces $H_1=0$ and $H_2=0$ do not
coincide, as the transformations (\ref{bs}) leave them intact.
On the other hand, the surfaces $H_{1,2}=0$ are just the ones the particle
dynamics is concentrated on (the constraint surfaces), therefore, one may
say that the particle dynamics fills the invariance w.r.t. hyperWeyl
transformations (\ref{weyl}). Besides, due to the hyperWeyl symmetry the
unphysical Hamiltonians which have no zeroes are equivalent to a constant.
Thus, at least infinitesimal equivalence transformations do not identify
all hamiltonians.

Still, the simple comment
that any two Hamiltonians are connected by a (rather singular) canonical map
has hard consequences for dynamical content of a theory
we will try to construct in terms of $H$.
The final result will be that in order to allow a nontrivial dynamics for
$H$ the generalized equivalence principle should be
either modified or relaxed, otherwise no sensible dynamical theory
constructed in terms of $H$ exists.

\section{An example: gravity+electromagnetic+scalar background.} {\label{low}}

Our consideration has been general till this moment. Let us consider
the important example to illustrate that our definition
(\ref{def}) of equivalence has a direct physical interpretation.

We will say that a function $f(x,p)$ is of a $k$-th order, if
it possesses the following decomposition in momenta:
\be \label{hclass0}
f = \sum\limits_{i=0}^{k} f^{m_1...m_i} (x) p_{m_1}...p_{m_i}
\ee
The example deals with the {\it second order} Hamiltonians
\be  \label{hamquadr}
H=D(x) + C^m(x)p_m + \frac{1}{2}g^{mn}(x) p_m p_n,
\ee
It is easy to see that these Hamiltonians describe a point
particle in a general gravity-electromagnetism-dilaton
background\footnote{In this paper, "dilaton" stands just for a scalar
field, with no other limitations on its dynamics, like the connection to
string theory etc.}. Indeed, the lagrangian action of such a particle
with nonzero mass is
\be  \label{lagpart}
S=\int d\tau \{-m\phi(x)\sqrt{-g_{mn}\dot{x}^m \dot{x}^n } + eA_m \dot{x^m}\}
\ee
where $g_{mn}(x), A_m (x), \phi(x)$ are the gravitational,
electromagnetic and dilaton fields, and the parameters $m$ and $e$ are the
particle's mass and electric charge, respectively. Carrying out the
hamiltonization, one finds the first class constraint
\be  \label{ham2}
H=\frac {1}{2} g^{mn}(x) \Pi_m \Pi_n +
\frac{m^2}{2} \phi^2(x),\qquad \Pi_m =p_m -eA_m
(x), \ee
$g^{mn}$ is an inverse metric, $p_m$ are the momenta and $\Pi_m$ are
the extended momenta. Now it is seen that
a general second order Hamiltonian (\ref{hamquadr})
is equal to (\ref{ham2}) after the identification
\be  \label{ham22}
D=\frac{m^2}{2} \phi^2 +\frac{e^2}{2} g^{mn} A_m A_n, \qquad C^n =-e\, g^{mn} A_n.
\ee
In fact, the Hamiltonian description is more general than the one
via "square root" Lagrangian \re{lagpart}, as it includes
massless limit ($m=0$ case).

Consider the equivalence transformations (\ref{bs})
which do not change the second order of
Hamiltonians and therefore may be
interpreted as the equivalence transformations for the Hamiltonian
coefficients, i.e. for metric, electromagnetic potential and dilaton. To
this end, it is sufficient to make the following choice:
\be \label{bs2}
\e(x,p) = \ve(x) +\xi^m (x) p_m\;\;,\;\; a(x,p)= \alpha(x),
\ee
so that the generator of canonical transformations $\e(x,p)$ is
taken to be a general {\it first  order}
function  while $a(x,p)$ does
not depend on momenta at all ({\it zero order}).
The transformations (\ref{bs},\ref{bs2})
form the infinite-dimensional Lie algebra  ${\cal G}_0$
\be \label{algsym2}
\begin{array}{c}
[\delta_{(\e_1,\al_1)} , \delta_{(\e_2,\al_2)} ] H =  \delta_{(\e_3,\al_3)}  H\\ \\
\ve_3= -\xi^m_1 \pa_m \ve_2 +\xi^m_2 \pa_m \ve_1\;,\;
\al_3= -\xi^m_1 \pa_m \al_2 +\xi^m_2 \pa_m \al_1\;,\;
\xi^n_3= -\xi^m_1 \pa_m \xi^n_2 +\xi^m_2 \pa_m \xi^n_1
\end{array}
\ee
which is unambiguously interpreted as semidirect sum of spacetime
diffeomorphisms $\xi_m$ and two $U(1)$ gauge symmetries corresponding to
the phase canonical transformations $\ve(x)$ and ordinary Weyl
dilations $\al(x)$.

It is easy to see that (\ref{bs},\ref{bs2}) give the following {\it gauge
transformations} for the background fields:

\be \label{glaws}
\begin{array}{c}
\delta g_{mn} =-\xi^k g_{mn,k} - {\xi^k}_{,m} g_{kn} - {\xi^k}_{,n}
g_{km} -\al g_{mn}\, \\ \\ \delta \,
e\,A_m = - e\,(\xi^k A_{m,k} +{\xi^k}_{,m} A_k) +
\ve_{,m} \,\\ \\ \delta \phi = -\xi^k \phi_{,k} +\frac{1}{2}\al \phi\end{array}
\ee
Here one  easily recognizes again the standard $U(1)$ and
general-coordinate transformations associated with $\ve$ and $\xi^m$,
correspondingly, and the Weyl dilations
associated with $\alpha$. Therefore, the
{\it equivalence} transformations (\ref{bs}) are interpreted as
{\it gauge} ones for the coefficients of Hamiltonian.
It is clear that the gauge equivalent backgrounds lead to the particle
dynamics which are physically equivalent. Indeed,
the invariance related to $\alpha$ is automatically accounted for
in the Lagrangian (\ref{lagpart}) as all fields
appear in the invariant combinations $\phi^2 g_{mn}, A_m$.
It is also well known that the Lagrangian (\ref{lagpart})
is covariant w.r.t. general coordinate
transformations of background fields and changes by total derivative
if electromagnetic field transforms under $U(1)$ symmetry.
It means that  gauge transformations of metric, electromagnetic field
and dilaton may be {\it compensated} by transferring to new coordinates
in the space of worldlines ${x'}^m(\tau)(x^n(\tau))$. Needless to say, this
map is induced by the canonical transformation with generating function
$\e(x,p)$ \re{bs2}.

Given the gauge laws \re{glaws} one may wonder if there exists an invariant
action
${\cal A}[g_{mn},A_m,\phi]$. Clearly the answer is positive and is given by
the action of Weyl-invariant dilaton-Maxwell gravity
\be \label{s2}
\ba
{\cal A}[g_{mn}, A_m, \phi] =\\ \\
= \frac{1}{4} \int d^d x \sqrt{-det( g)} \phi^d \;
\bigl\{ \Lambda  + a_1( (d+2)^2 g^{pq} \phi_{,p} \phi_{,q} +\frac{d-2}{d-1} R(g)\phi^2) +
a_2 F_{pq} F^{pq}\phi^{-4}  +... \bigr\}
\ea
\ee
where $a_{1,2}$ are some constants, $\Lambda$ is a cosmological constant, $F_{mn}$ is the electromagnetic field strength,
$R$ is the scalar
curvature and "$...$" stand for higher derivative terms.
If one drops the higher derivative terms, the action describes
the dynamics equivalent to ordinary Einstein gravity (possibly, with
cosmological term) plus Maxwell theory. This can be seen by
gauging the dilaton to a nonzero constant by Weyl transformations.
The special singular case arises when the dilaton is allowed to take zero value.
The most simple way to deal with this limiting value  seems to
be the redefinition of the dilaton according to the rule
\be
\Phi = \phi^{\frac{d+2}{2}}
\ee
to get
\be \label{s22}
\ba
{\cal A}[g_{mn}, A_m, \Phi] =\\ \\
= \int d^d x \sqrt{-det( g)} \;
\bigl\{ \Lambda \Phi^{\frac{2d}{d+2}} + a_1( g^{pq} \Phi_{,p} \Phi_{,q} +
\frac{1}{4} \frac{d-2}{d-1} R(g)\Phi^2) +
a_2 F_{pq} F^{pq}\Phi^{\frac{d-4}{d+2}}  +... \bigr\},
\ea
\ee
wherefrom it is seen that in the limit $\Phi \rightarrow 0$
only dilaton's kinetic term survives.

Let us summarize the results of this section:
the {\it second order} Hamiltonians describe scalar
particle in general gravity, electromagnetic and dilaton backgrounds.
The standard background fields gauge transformations: general coordinate,
$U(1)$ and Weyl dilations, do not affect the particle dynamics in the
sense they do not change the action after compensation by a
canonical transformation with generating function of a
{\it first order}.
Needless to say, these properties are nothing but a paraphrase of
the equivalence principle.
There exists the standard action \re{s2} invariant w.r.t. equivalence
transformations of background fields which appear to describe (at general
values of dilaton) Maxwell $+$ Einstein gravity system.
As far as the form of the action \re{s2} is known to be fixed by the
requirements of gauge invariance \re{glaws}
and absence of higher derivatives
one may conclude that the equivalence principle applied to the
second order ansatz \re{hamquadr},\re{bs2} determines uniquely
the  theory action and thus
sets up the whole Einstein $+$ Maxwell theory.

On the other hand the
standard low-spin gauge fields saturates only quadratic ansatz in momenta
\re{hamquadr},\re{ham22},
while it is clear
one will get a consistent particle dynamics on ${\cal M}^d$ for more
general Hamiltonians than the quadratic ones.
Therefore, the generalized equivalence principle will work for general case
either and it may happen to determine some interesting gauge theory with infinite number
of gauge fields and gauge invariances.
This possibility
is studied in the next section.

\section{General backgrounds}\label{fff}

The facts exposed in the previous section
are well known. Point out once again the affinity between the geometry of point
particle phase space and the geometry of gauge fields -- in fact the
full (nonlinearized, background-independent) general coordinate and $U(1)$
transformations of gravity and electromagnetism are induced by the
canonical transformations of the scalar particle phase space.

Here we extend these observations by working out general
(in an appropriate class of functions)
Hamiltonians  and general equivalence transformations. It will
provide us with a huge algebra of gauge transformations acting on the
coefficients of Hamiltonian identified with an infinite set of
space-time fields.
Proceeding in this way one gets
the gauge transformation laws for the infinite system of
symmetric tensor fields.
The infinitesimal gauge transformations
(\ref{bs}) read
\be  \label{bs7}
\delta H(x,p)  =  \{\e(x,p), H(x,p) \} +  a(x,p) H(x,p)
\ee
Consider the Hamiltonians which are {\it the formal series in momenta}
\be \label{hclass}
H = \sum\limits_{k=0}^\infty H^{m_1...m_k} (x) p_{m_1}...p_{m_k} =
\sum\limits_{k=0}^\infty H_k
\ee
where $H_k$ denotes the polynomial of $k$-th degree in momenta.
Similarly, the equivalence transformations (\ref{bs}) are generated
by the parameters $\e(x,p)$ and $a(x,p)$, which belong to the same class
\be  \label{eaclass}
\e = \sum\limits_{k=0}^\infty \e^{m_1...m_k} (x)
p_{m_1}...p_{m_k}\;\;,\;\; a = \sum\limits_{k=0}^\infty a^{m_1...m_k} (x)
p_{m_1}...p_{m_k}
\ee
and required to have a compact support in space-time.
$x$-diffeomorphisms, $U(1)$ and Weyl dilations
(\ref{bs2},\ref{algsym2}) form the subalgebra ${\cal G}_0$.
Note that the full gauge algebra
${\cal G}$ gets broken to ${\cal G}_0$ if one restricts
the sums in (\ref{hclass},\ref{eaclass}) by some highest $k$.
Therefore, the infinite number of component fields $H^{m_1...m_k}$ is
necessary to have not only ${\cal G}_0$ but also its infinite extension.
This property naturally corresponds to the requirement (2) of
infinite number of higher spin fields mentioned in the Introduction
among the conditions necessary for consistent interactions.

The coefficients $H^{m_1...m_k}(x)$ are interpreted as background
space-time gauge fields. One may set the question whether
it is possible to construct an action ${\cal A}[H_0,H_1,...,H_k,...]$
invariant under the gauge transformations (\ref{bs7}).
Unfortunately,
these transformations appear to prohibit  the construction of
nontrivial invariant action functionals.

Indeed, the $\e$-variations of $H$ are ordinary canonical
transformations with an arbitrary generating function $\e$.
The
invariants of these transformations seem to be only the integrals
over total phase space
\be  \label{naiv}
I_F [H] = \int d^d p d^d x  F(H)
\ee
where $F$ is some function of $H$ such that the integral converges.

Furthermore
the hyperWeyl invariance with parameter $a$
implies that the function $F[H]$ should be dilation invariant:
\be
F(AH)=F(H)
\ee
for all $A(x,p) \neq 0$.
Then the unique possibility for $F(\sigma)$ is
\be \label{teta}
F(\sigma)= \theta(\sigma)
\ee
where $\theta(\sigma)$ is the  "step" $\theta$-function:
\be
\ba
\theta(\sigma)=0\;\;,\;\; \sigma < 0 \\ \\
\theta(\sigma)=\fr\;\;,\;\; \sigma = 0\\ \\
\theta(\sigma)=1\;\;,\;\; \sigma > 0.
\ea
\ee
Suppose that the surface $H=0$ is a boundary
of a compact domain $U$ in the phase space, and $H>0$ inside the domain while
$H<0$ outside, e.g. $H=\fr (-p^2 -\alpha^2 x^2 + m^2)$ with Euclidean metric.
Then the action (\ref{naiv},\ref{teta})
equals $\fr$ times the volume of this compact domain ($2n-1$ dimensional sphere
in the last example).
An analogous situation arises when the surface $H=0$ contains
a number of nested disconnected pieces (e.g. $H= \prod \limits_{i=1}^{N} H_i\;,\;
H_i=-p^2 -\alpha x^2 +m_i^2\;,\;m_1< m_2 < ... < m_N$), then
the volume between $H_i$ and $H_j$ surfaces is invariant w.r.t.
(\ref{bs7}) and may be taken as an invariant "action".

Clearly, such type of actions is inappropriate
as they {\it do not contain space-time derivatives of dynamical fields} and
therefore do not set a nontrivial dynamics.
One may wonder why the full gauge transformations are too restrictive while
for the quadratic ansatz of the previous section the equivalence transformations
provide standard gauge laws for gravity+Maxwell+dilaton fields which
are known to provide good space-time actions.

A possible answer which we do not dwell here is that the equivalence transformations
should be deformed in order to allow for  actions with derivatives.
In fact, there exist a very natural deformation related to the
{\it covariant quantization} of the point particle in general background fields,
in that case one easily constructs the actions of the type
\be \label{quantiz}
{\cal A}_H =Tr F(\hat{H}\left(\hat{x},\hat{p})\right),
\ee
where $\hat{H}$ is a quantized
Hamiltonian and the trace is performed over the quantized particle states
space. It is possible to show that the functionals of this kind possess
natural {\it quasiclassical} expansion by the powers of $\hbar$, and
the $\hbar^k$-order terms have exactly $k$ space-time derivatives of
the $\hat{H}$ components while $\hbar^0$-terms does not contain derivatives
and coincides with the classical result \re{naiv}. So, in this scheme, the
terms with derivatives appear as {\it quantum corrections} to the classical
"cosmological" term.
We plan to present these matters in a separate publication.

Here we try to give another answer which makes it evident
from the physical viewpoint that
the generalized equivalence transformations are too restrictive.
To provide the evidence we have to analyze the linearization of
transformations \re{bs7}. We expand the Hamiltonian  $H(x,p)$
around a natural background,
\be  \label{vacuum}
H={H_v}  =
\frac{1}{2} (g^{mn} p_m p_n +m^2 )\equiv
\frac{1}{2} ( p^2 +m^2).
\ee
which presents a configuration with nonzero metric, constant dilaton
and all other fields being zeroes.
Let us suppose there exists an action $\tilde{{\cal A}}[H]$ which possesses
the vacuum solution $H_v$ and describes well-defined space-time dynamics
of $H$. Introduce the fluctuation $h$ near the vacuum   ${H_v}$,
\be
H= {H_v}  + h.
\ee
and extract the $h$-quadratic part from $\tilde{{\cal A}}[H]$, i.e.
construct the linearized action $\tilde{{\cal A}}_2[H]=hPh$, where
$P$ is some gauge-invariant operator.

Recall a general proposition:
if one has a nonlinear action ($\tilde{{\cal A}}[H]$) with
infinitesimal gauge invariance $\delta H = R_{\varepsilon}[H]$,
then the linearized action $hPh$ is gauge invariant w.r.t.
linearized inhomogeneous, $h$-independent gauge
transformations given by gauge variation of the vacuum solution: $\delta
h = R_{\varepsilon}[{H_v}]$.
We suppose  the gauge invariance of the action
differs from \re{bs7} only at interaction level while the difference
may be neglected in the linearized approximation.
Then the linearization of gauge transformations \re{bs7} reads
\be  \label{bs4}
\delta h  =  \{\epsilon, {H_v} \} +  a {H_v},
\ee
Note that for  $h$ satisfying the
{\it second order ansatz} of the previous section \re{hamquadr}
with {\it first order } $\e$ and {\it zero order} $a$ \re{bs2}, these
gauge transformations reproduce the linearization of standard
general coordinate, $U(1)$ and Weyl symmetries \re{glaws} around the vacuum
\re{vacuum}:
\be \label{glaws1}
\begin{array}{c}
\delta h^{mn} =  {\xi^k}^{;m} + {\xi^k}^{;n}  +\fr \al g^{mn}\,
\\ \\ \delta h^m = \ve^{,m} \,\\ \\
\delta h = \frac{1}{2}\al m^2\end{array}
\ee
where "$;$" denotes covariant differentiation w.r.t. vacuum metric.
For $m^2 \neq 0$ these transformations  imply that the
invariant quadratic actions should be the standard
spin-1 and spin-2 ones. The "spin-0" fluctuation of dilaton $h$ is gauged away
by linearized Weyl transformations. Another way to see it is to observe that due to the
Weyl symmetry $h$ combines with the trace of metric fluctuation ${h^m}_m$
into the single invariant combination
\be
h_{m^2}= h -\frac{m^2}{d} {h^m}_m
\ee
which then merely serves as the counterpart of another Weyl-invariant object:
traceless part of $h^{mn}$ in building the standard linearized
gravity action.  So the spin-$0$ is absent here and this is in the complete
agreement with nonlinear situation discussed in the previous section.

If $m^2=0$ the fluctuation of dilaton $h$ is gauge
invariant and therefore it decouples from the linearized action
describing  fluctuations of rank 1 and 2 fields. On the other hand,
the Weyl transformations in gravity fluctuations sector gauge away the
trace ${h^m}_m$ thus the invariant action has to depend only on
the traceless part of $h^{mn}$. It results in Weyl-invariant
higher derivative gravity theories without zero and second-order terms
(e.g. in $4D$ the Lagrangian is the linearization of
$\sqrt{-g}\, C_{mnpq}^{\,2}$ for $C_{mnpq}$ being the Weyl tensor) $+$
further higher derivatives terms. One concludes that
the fluctuations around the $m^2=0$ vacuum are nonunitary and
thus this case does not provide a possibility to interpret the fluctuations
in terms of particles.

Let us study the general case \re{bs4}
and try to find a relationship of general fluctuations
around the vacuum $H_v$ to the Fronsdal double-traceless gauge fields
\re{dtrace},\re{frlaw}.
Despite the overall result will be {\it negative}
both for $m^2\neq 0$ and $m^2=0$ cases it will be seen
how the fluctuations of Fronsdal fields can be embedded
into general fluctuations of Hamiltonian $h(x,p)$.
For simplicity, we take the vacuum metric to be flat.

Let us start with the gauge invariance
(\ref{vacuum},\ref{bs4}) which reads
\be \label{bsactual}
\delta h = \fr a (p^2 +m^2) + p^m \pa_m \e,
\ee
where $\pa_m$ is a derivative w.r.t. $x^m$.
The linearized hyperWeyl transformations parameters $a(x,p)$ enter the gauge laws \re{bsactual}
without  derivatives and therefore they just eliminate some auxiliary fields.
If an invariant action $\tilde{{\cal A}}_2[H]=hPh$ exists, the
invariance w.r.t. $a$ implies these
auxiliary fields do not enter action at all and the action depends only on
the invariants of $a$-transformations.
So let us find these invariants. It can be done as follows.
Given an arbitrary power series in momenta
$f$ of the form \re{hclass}
one may expand each coefficient
$f^{m_1...m_i}$ in terms of its traceless components in the manner
\be
f^{m_1...m_i}= f^{m_1...m_i}_0 +g^{(m_1 m_2} f^{m_3...m_i)}_1 +
g^{(m_1 m_2} g^{m_3 m_4} f^{m_5...m_i)}_2 +...,
\ee
where all  $f_k$ are traceless, so any
$f$ may be represented as
\be  \label{decom}
f= f_0+ f_1 \,p^2 +f_2\, (p^2)^2 +... = \sum \limits_{k=0}^{\infty} f_k \,(p^2)^k,
\ee
where $f_k$ are the power series of the form \re{hclass},
but with {\it traceless}
coefficients. The expansion \re{decom} may be considered as a function $f(p^2)$
of one
variable $p^2$ with coefficients $f_k$ taking values in the
power series \re{hclass} with traceless coefficients,
then \re{decom} provides the Taylor
series of the function $f(p^2)$ at the point $p^2=0$.
Decomposing the function at another point,
say, $p^2=-m^2$, one gets
\be  \label{decom2}
f= f^{(m^2)}_{0}+ f^{(m^2)}_{1} \,(p^2 +m^2) +f^{(m^2)}_{2} (p^2 +m^2)^2 +...
= \sum \limits_{k=0}^{\infty} f^{(m^2)}_{k} \,(p^2+m^2)^k
\ee
Then it is clear
the unique invariant of hyperWeyl transformations \re{bsactual} is given by
\be \label{trick}
h_{m^2}\equiv h^{(m^2)}_{0} =\sum \limits_{k=0}^{\infty} (-m^2)^k h_k ,
\ee
or, formally, $h_{m^2} =h|_{p^2 +m^2 =0}$, where the reduction to the
"constraint surface" $p^2+m^2=0$ is carried out only in the $p^2$-factors
of \re{decom}. Indeed, representing $h$ and the hyperWeyl variation of $h$ \re{bsactual}
in the form \re{decom2} one observes that $a$ gauges away all $h^{(m^2)}_{k}$
components except $h^{(m^2)}_{0}$.

$h_{m^2}$ is a power series comprising only {\it traceless}
coefficients
\be
h_{m^2} =
\sum \limits_{k=0}^{\infty} \chi^{a_1 ... a_k} p_{a_1} ...p_{a_k} \;\;;\;\;
{\chi_b}^{b a_3 ... a_k} =0.
\ee
Now analyze how the linearized canonical transformations $\e$ \re{bsactual}
act on $h_{m^2}$. To this end one decomposes $p\pa_x\e$
in power series \re{decom2} and then substitutes $-m^2$ instead of $p^2$.
The parameter $\e$ may be taken traceless
\be
\e =\e^{(m^2)}_0 =
\sum \limits_{k=0}^{\infty} \ve^{a_1 ... a_k} p_{a_1} ...p_{a_k} \;\;;\;\;
{\ve_b}^{b a_3 ... a_k} =0
\ee
as its traces $\e=(p^2 +m^2)\e^{(m^2)}_1 +...$
lead to transformations equivalent to hyperWeyl ones and
already projected out automatically in the variation of $h_{m^2}$.
One easily finds
\be
\ba
\delta h_{m^2}=(p\pa_x \e)(-m^2) =\\ \\
\sum \limits_{k=0}^{\infty} \;\left(\;\pa^{a_1}\; \ve^{a_2 ... a_k} -
\frac{k-1}{d+2k-4} \;g^{a_1a_2}\; \pa_b \;\ve^{b a_3... a_k} -
m^2 \frac{k+1}{d+2k} \;\pa_b \;\ve^{b a_1 a_2 ...a_k}\;\right) \;p_{a_1} p_{a_2}...p_{a_k}
\ea
\ee
or equivalently
\be \label{corollary}
\ba
\delta
\chi^{a_1 ... a_k} =
\mbox{Traceless part of}\; \pa^{(a_1}\; \ve^{a_2 ... a_k)} -
m^2 \frac{k+1}{d+2k} \;\pa_b \;\ve^{b a_1 a_2 ...a_k}
\ea
\ee
where the round brackets denote the symmetrization.

For $m^2 \neq 0$, each tensor of rank $k$ is
transformed by two different gauge parameters: of rank $(k-1$) via
the "Traceless part of ..." in \re{corollary} and of rank
$(k+1)$ via the divergence-like term in \re{corollary}.

Note that restricting \re{corollary} to
the sector of $\chi, \chi^a, \chi^{a_1 a_2}$ and choosing the situation
with the only nonzero parameters $\ve, \ve^a$ one finds
for $m^2 \neq 0$ exactly
the linearized gauge transformations for Maxwell+Einstein system, where
$\chi, \chi^{a_1 a_2}$ describe the off-shell graviton and
$\chi^a$ the off-shell photon.
Thus we come back to the low-spin analysis carried out above
which showed how gravity +Maxwell fields arise by virtue of
the equivalence principle.
However, the higher rank gauge symmetries break good low-spin picture.
In particular, the graviton transforms not only w.r.t. linearized
diffeomorphisms $\ve^a$ but also w.r.t. $\ve^{a_1 a_2 a_3}$:
\be \label{corollary1}
\ba
\delta
\chi^{a_1 a_2} =
\mbox{Traceless part of}\; \pa^{(a_1}\; \ve^{a_2)} -
\frac{3m^2}{d+4} \;\pa_b \;\ve^{b a_1 a_2}
\ea
\ee
So,
despite the analogy with the low-spin case
was the main motivation for the consideration of generalized equivalence
principle, the results of its promotion to general case \re{corollary}
are not very appealing. It may be traced back to the
simple fact noted above that, from the very beginning,
the number of canonical  gauge parameters $\e$
coincides with the number of $H$ components and therefore
any two Hamiltonians  are identified by a singular canonical transformation.

On the other hand, let us compare the gauge laws \re{corollary} with
those of Fronsdal theories \re{dtrace} -- \re{fraction}.
Recall the Fronsdal gauge transformations for
true higher spin fields \re{dtrace} -- \re{etrace}. Decomposing
$\vp^{a_1...a_k}$ as a sum of its traceless part $\rho^{a_1...a_k}$ and
the trace
$\sigma^{a_1...a_{k-2}}$
\be
\vp^{a_1...a_k} =\rho^{a_1...a_k} + g^{(a_1 a_2}
\sigma^{a_3...a_{k})}
\ee
one gets
\be
\delta  \rho^{a_1...a_k} =
\mbox{Traceless part of}\; \pa^{(a_1}\; \ve^{a_2 ... a_k)} \;\;,\;\;
\delta
\sigma^{a_3...a_{k}}=\;\frac{k-1}{d+2k-4}\; \pa_b \ve^{b a_3...a_k}.
\ee
In Fronsdal theories, the "physical" field describing the dynamics
is
$\rho^{a_1...a_k}$ while
the trace $\sigma^{a_1...a_{k-2}}$ serves as a compensator designed to
ensure the constraint $\pa_b\ve^{b a_3...a_k}=0$ after $\sigma$
is eliminated by gauge transformations.
Comparing this situation to \re{corollary},\re{corollary1}
one concludes that each traceless component
$\chi^{a_1...a_k}$ plays the {\it double role}
: first, it transforms as "physical" field $\rho^{a_1...a_k}$
w.r.t. $\ve^{a_1...a_{k-1}}$, second, it transforms like the
compensator $\sigma^{a_1...a_k}$ w.r.t. $\ve^{a_1...a_{k+1}}$.

As a consequence, the gauge laws \re{corollary} do not determine
any sensible free theory.
Possibly, the situation with the gauge transformations \re{corollary}
may be improved if one relaxes the generalized equivalence principle,
e.g. by relaxing somehow hyperWeyl transformations or finding
some constraints on $\e$ and $a$ \re{bs} which
restrict the gauge transformations in such a way
they allow  nontrivial dynamics for $H$. Still, this
modification has to possess good features of generalized equivalence principle:
background independence, inclusion of general coordinate, $U(1)$ and
higher gauge symmetries, clear physical interpretation etc.

On the other hand, our study brings
the idea of {\it doubling}: if one replaces the fluctuation $\chi$
by two fields $\chi_1, \chi_2$ transforming by the laws \re{corollary}
with two {\it different} mass values
\be \label{corollary2}
\ba
\delta
\chi_1^{a_1 ... a_k} =
\mbox{Traceless part of}\; \pa^{(a_1}\; \ve^{a_2 ... a_k)} -
m_1^2 \frac{k+1}{d+2k} \;\pa_b \;\ve^{b a_1 a_2 ...a_k} \\ \\
\delta
\chi_2^{a_1 ... a_k} =
\mbox{Traceless part of}\; \pa^{(a_1}\; \ve^{a_2 ... a_k)} -
m_2^2 \frac{k+1}{d+2k} \;\pa_b \;\ve^{b a_1 a_2 ...a_k},
\ea
\ee
then these gauge laws determine the system exactly equivalent to the
infinite collection of Fronsdal fields, every spin enters once, where
the "physical" $\rho$ and compensator $\sigma$ components of Fronsdal
fields are expressed via linear combinations of $\chi_{1,2}$.
The derivation of the system \re{corollary2} as a linearization of some
sensible nonlinear  transformations analogous to the
derivation of \re{corollary} from \re{bs} is an interesting issue  we plan to
study in a separate paper.

Now let $m^2 =0$. Then the gauge laws \re{corollary} simplify:
\be \label{corollary3}
\delta
\chi^{a_1 ... a_k} =
\mbox{Traceless part of}\; \pa^{(a_1}\; \ve^{a_2 ... a_k)}.
\ee
Here $\chi$ transforms as "physical" Fronsdal field
and the low spin situation is not broken by higher spin gauge symmetries.
However, the "compensators" are absent and therefore the gauge laws
\re{corollary3} has no relation to Fronsdal theories.
The low spin case was shown above to lead, in $4D$, to the
linearization of (higher derivative)
conformal gravity.
There may exist some nonunitary higher derivative actions invariant w.r.t.
the symmetry \re{corollary3}
which provide the "higher spin" generalizations of
linearized conformal gravity.

\section{De Wit - Freedman point particle - symmetric tensors first
order interaction. } \label{DF}

Concluding the analysis of classical particle model
it is worth mentioning  that our approach allows to reproduce
and generalize some old results.
Recall that an attempt to find a point particle-symmetric tensor fields
interaction in the lagrangian approach
had been undertaken in \cite{DeWitFreedman} by De Wit and Freedman who
succeeded in constructing  consistent first-order interaction
linear in background fields fluctuations.
Here we show that their results are {\it identically} reproduced if one
extracts linear in the fluctuations of external fields terms from the
Lagrangian derived from the full action
\re{ham}, while our construction appears to generalize
the De Wit - Freedman results to all orders in fluctuations.

Consider the full hamiltonian action \re{ham}. Represent the Hamiltonian in the manner
\be
H=\fr p^2 +\fr m^2 +eh(x,p)
\ee
where $\fr (p^2 + m^2) $ corresponds to the vacuum configuration of
background fields (with arbitrary metric) and $eh$ is a fluctuation over
this vacuum with $e$ being the expansion parameter. The action takes the form
\be \label {hamBdW}
S_H [x(\tau), p(\tau), \lambda(\tau)]=\int d\tau \{p_m \dot{x}^m -\lambda
\left(\fr p^2 +\fr m^2 +eh(x,p)\right)\}.
\ee
In accordance with \re{invar} this action is invariant under the gauge
transformations
\be  \label{invarBdW}
\delta{x^m(\tau)}= (p^m + e \frac{\pa}{\pa p_m} h) \mu(\tau), \qquad
\delta{p_m(\tau)}=- e (\frac{\pa}{\pa x^m} h) \mu(\tau),
\qquad \delta \lambda=-\dot{\mu}.
\ee
Let us pass to the Lagrangian formalism. To this end, one has to exclude
the momenta $p_m$ by means of their equations of motion
\be \label{moment}
\frac{\delta S}{\delta p_m} = \dot{x}^m -\la (p^m + e \frac{\pa}{\pa p_m} h )=0.
\ee
This algebraic (w.r.t. $p_m$) equation is solved by iterations in $e$. For our purpose it is
sufficient to get the solution up to $o(e^2)$ terms.
Expanding $p_m$ in the manner
\be \label{pexpan}
p_m =p_m^{(0)} + e p_m^{(1)} +e^2 p_m^{(2)}+...
\ee
one obtains
\be  \label{pexpan1}
p_m^{(0)} = \la^{-1} \dot{x}^m\;\;,\;\;
p_m^{(1)} =- \frac{\pa}{\pa p_m} h (p \rightarrow \la^{-1} \dot{x}),
\ee
wherefrom it follows that
\be
\ba
p\dot{x} =\la^{-1} \dot{x}^2 - e\la p\pa_p h (\la^{-1} \dot{x})  + o(e^2)\\ \\
p^2=\la^{-2} \dot{x}^2 - 2 e\la p\pa_p h (\la^{-1} \dot{x})  + o(e^2)\\ \\
eh= e h (\la^{-1} \dot{x})  + o(e^2).
\ea
\ee
Substituting these expressions into  \re{hamBdW} we obtain the
lagrangian action
\be \label {lagBdW}
S_L[x(\tau),\la(\tau)]
=\int d\tau \{\fr (\la^{-1} \dot{x}^2 -m^2 \la)-e\la h (\la^{-1} \dot{x})\}
+o(e^2 \,h^2).
\ee
The $e$-independent terms give the well known scalar particle action in the
gravitational background, while $e$-linear ones present higher rank
corrections. To make contact with De Wit -Freedman Lagrangian one excludes
$\la$ by means of it's equations of motions
\be \label{la}
\frac{\delta S_L}{\delta \la} =
-\fr m^2-\fr \la^{-2} \dot{x}^2  -
e \pa_{\la} h (\la^{-1} \dot{x}) +o(e^2)=0.
\ee
Like in the case of momenta, these equations are solved by iterations in
$e$.
Expanding $\la^{-1}\equiv y$ in the manner
\be  \label{laexpan}
y =y_{(0)} + e y_{(1)} +e^2 y_{(2)}+...
\ee
one gets
\be
\ba  \label{laexpan1}
y_{(0)} =[-\frac{m^2}{\dot{x}^2}]^{\fr} \\ \\
y_{(1)} =  m^{-2}
\sum \limits_{k=0}^{\infty} (1-k)
h_{m_1...m_k}\dot{x}^{m_1}...\dot{x}^{m_k}
(-\frac{m^2}{\dot{x}^2})^{\frac{k+1}{2}}.
\ea
\ee
In fact, the explicit form of $y_1$ is inessential as in our first order
approximation its contribution drops out from the action.
Substituting \re{laexpan1} in  \re{lagBdW} we arrive at the result
\be \label {hamBdW1}
-  S_L[x(\tau)]
=\int d\tau
\{ \sqrt{-m^2\dot{x}^2} \left(1+ \frac{e}{m^2}
\sum \limits_{k=0}^{\infty}
h_{m_1...m_k}\dot{x}^{m_1}...\dot{x}^{m_k}
(-\frac{m^2}{\dot{x}^2})^{\frac{k}{2}} \right)\}
+o(e^2 h^2).
\ee
This action coincides with the De Wit-Freedman (DF) one
\cite{DeWitFreedman} after the identification
$\varphi^{DF}_{m_1...m_k} = - h_{m_1...m_k}$ and
setting $m^2=-1$ (the negative sign of $m^2$
just accounts the difference in metric's signature,
$\sqrt{\dot{x}^2}{}|_{DF}=\sqrt{-\dot{x}^2}|_{our}$).

In their work it was observed that, besides being explicitly
reparametrization invariant, this lagrangian action is invariant
w.r.t. simultaneous gauge transformation of external fields
\be \label{DFtrs0}
\delta \varphi^{DF}_{m_1...m_k} = k \xi_{(m_2...m_k ; m_1)}
\ee
($;$ denotes the covariant derivative) and point particle worldlines
\be \label{DFtrs}
\delta x^m (\tau)= -e k(k-1) (\dot{x^2})^{1-k/2}
{\xi^m}_{m_3...m_k}  \dot{x}^{m_3}...\dot{x}^{m_k}
\ee
In our formulation, this invariance is a particular manifestation of the
generalized equivalence principle \re{Equivalence}, corresponding to the
canonical equivalence transformations \re{canon}
written up to $o(e^2)$ terms.
Explicitly, writing down the infinitesimal canonical $\e$-transformations
(\ref{canon},\ref{bs})
\be   \label{DFequiv}
\ba
\delta x^m =-e\,\{\e, x^m\}\\ \\
\delta H =e\,\{\e, H\} \Rightarrow \delta\; h =\fr \, \{\e, p^2\} + o(eh)
\ea
\ee
and substituting the expressions for momenta and Lagrange multipliers
(\ref{pexpan},\ref{pexpan1},\ref{laexpan},\ref{laexpan1}), one gets
in the $o(e^2)$ approximation the
De Wit-Freedman transformations \re{DFtrs0},\re{DFtrs}
after the identification
$\e_{m_2...m_k}= - k \xi_{m_2...m_k}$, while the invariance of the action is guaranteed
by the very definition \re{Equivalence}.

Doing further orders of perturbation procedure for Eqs.
\re{moment},\re{la} and substituting the solutions
for
$p_m$ and $\la$ into the action
\re{hamBdW} one obtains the Lagrangian up to an arbitrary order in
$e\,h$. Then the generalized equivalence transformations
\re{DFequiv} present the generalization
of \re{DFtrs0},\re{DFtrs} to all orders in $eh$.

\section{Conclusion}

We have considered the model of point particle in general
external fields and postulated the generalized equivalence
principle (Definition \re{def}).
This principle identifies background fields which set up
equivalent particle dynamics, with constraint surfaces $H(x,p)=0$
differing by a canonical transformation.
The Hamiltonians $H(x,p)$ have a form of formal
power series in momenta $p$ and coefficients of power series are identified
with an infinite collection of space-time symmetric tensors.

It is shown that {\it second order} Hamiltonians which saturate
the {\it quadratic} ansatz in momenta correspond to general
{\it gravitational+Maxwell+dilaton} background. In this case, the equivalence
transformations which do not spoil the ansatz give exactly the
general coordinate, $U(1)$ and Weyl transformations.

The generalization to arbitrary Hamiltonians and
equivalence transformations is worked out to some depth.
The general equivalence transformations appear to prohibit the construction of
invariant actions with space-time-derivatives.
It is not too surprising as the equivalence transformations
contain as many parameters as Hamiltonian.
The invariants of full equivalence transformations seem to be
only the integrals over the total phase space
\re{naiv} which do not contain space-time derivatives of $H(x,p)$ and
therefore do not set a nontrivial dynamics.
This defect could be cured after quantizing the particle and considering
the functionals of the form \re{quantiz}, we plan to consider this possibility in a separate paper.

On the other hand we tried to see explicitly why
the generalized equivalence principle does not allow for good space-time actions while
the ordinary equivalence principle for low spin fields does.
To this end we have undertaken the analysis of fluctuations over a natural
(Minkowski) vacuum and found that the
surviving degrees of freedom are described by the infinite collection of traceless
symmetric  tensors (every rank enters once),
and linearized gauge transformations mix the components of all ranks.
Moreover, even low spin Maxwell+gravity system is broken as
graviton may be gauged away by gauge transformations with
third-rank gauge parameter. Comparing the situation with
the Fronsdal theory where a fixed spin-$s$ system
is described off-shell by two traceless fields -- rank-$s$ "physical" and rank
$(s-2)$ "compensator",
we observed that the linearized equivalence transformations
imply too hard duties on the fluctuations: every component of a fixed rank
$s$ transforms as a "physical" field w.r.t. rank-$(s-1)$ parameters
and as a "compensator" w.r.t. rank-$(s+1)$ ones.

Summing up, the generalized equivalence principle is too restrictive
and need to be relaxed either by finding appropriate constraints
on the canonical and hyperWeyl transformations or by modification
of the original gauge law \re{bs}. This hypothetical
relaxed equivalence principle has to possess background independence,
contain general-coordinate and $U(1)$ transformations
as well as infinity of higher spin symmetries, and has a clear physical
interpretation.

On the other hand, the situation could be improved after introducing some additional degrees of freedom
which should have some clear physical origin and may be
incorporated naturally in the framework of generalized equivalence.
Particularly interesting is the "doubled" linearized system \re{corollary2} which
is exactly equivalent to the infinite collection of Fronsdal gauge fields,
and the question is
whether this system may be obtained by linearization
from some background independent gauge transformations, in that case
the full transformations could present
the nonlinear higher spin gauge symmetries.
We leave this topic for future study.

We also have shown that the De Wit -- Freedman first order point particle-
symmetric tensors lagrangian interaction \cite{DeWitFreedman}
is reproduced naturally in our approach  if one
derives it from the Hamiltonian action \re{ham}  excluding the momenta
and Lagrange multiplier by means of their equations of motion in the
framework of simple perturbative procedure, this way the
invariance of the action w.r.t. simultaneous gauge transformations
of background fields and worldlines of the particle found in \cite{DeWitFreedman}
arises as a particular
manifestation of generalized equivalence principle (which is satisfied
by construction in all orders of perturbative procedure).

We hope our observations may get important developments in the future.

\section{Acknowledgement} The author thanks
A.A. Sharapov for useful discussions.
The work is supported by grants RFBR 99-02-16207 and RFBR 00-15-96566.

\end{document}